\documentclass[showpacs, preprint, superscriptaddress, amsmath, amssymb, aps, prmaterials]{revtex4-2}

\usepackage{graphicx}
\usepackage{dcolumn}
\usepackage{bm}
\usepackage[utf8]{inputenc}
\usepackage{xfrac}
\usepackage[colorlinks=true, linkcolor=blue, urlcolor=blue, citecolor=blue]{hyperref}

\usepackage{booktabs}
\usepackage{makecell}

\usepackage{color}
\usepackage{xcolor}
\usepackage{siunitx}
\usepackage[version=4]{mhchem}
\usepackage{gensymb}
\emergencystretch=\maxdimen
\begin{document}
	
	\title{Superconductivity in epitaxial PtSb(0001) thin films}
	
	\author{\mbox{C. M\"uller}}
	\affiliation{Institute of Physics, Czech Academy of Sciences, 162 00 Prague, Czech Republic}
	\affiliation{Faculty of Mathematics and Physics, Charles University, 121 16 Prague, Czech Republic}
	
	\author{\mbox{S. P. Bommanaboyena}}
	\affiliation{Institute of Physics, Czech Academy of Sciences, 162 00 Prague, Czech Republic}
	
	\author{\mbox{A. Badura}}
	\affiliation{Institute of Physics, Czech Academy of Sciences, 162 00 Prague, Czech Republic}
	\affiliation{Faculty of Mathematics and Physics, Charles University, 121 16 Prague, Czech Republic}
	
	\author{\mbox{T. Uchimura}}
	\affiliation{Research Institute of Electrical Communication, Tohoku University, Sendai 980-8577 Japan}
	
	\author{\mbox{F. Husstedt}}
	\affiliation{Hochfeld-Magnetlabor Dresden (HLD-EMFL), Helmholtz-Zentrum Dresden-Rossendorf, 01328 Dresden, Germany}
	\affiliation{Institut für Festkörper- und Materialphysik, Technische Universität Dresden, 01062 Dresden, Germany}
	
	\author{\mbox{B. V. Schwarze}}
	\affiliation{Hochfeld-Magnetlabor Dresden (HLD-EMFL), Helmholtz-Zentrum Dresden-Rossendorf, 01328 Dresden, Germany}
	
	\author{\mbox{S. Banerjee}}
	\affiliation{Institute of Physics, Czech Academy of Sciences, 162 00 Prague, Czech Republic}
	\affiliation{Faculty of Nuclear Sciences and Physical Engineering, Czech Technical University in Prague, 120 00 Prague, Czech Republic}
	
	\author{\mbox{M. Ledinský}}
	\affiliation{Institute of Physics, Czech Academy of Sciences, 162 00 Prague, Czech Republic}
	
	\author{J. Michalicka}
	\affiliation{Central European Institute of Technology, Brno University of Technology, 612 00, Brno, Czech Republic} 
	
	\author{\mbox{M. Míšek}}
	\affiliation{Institute of Physics, Czech Academy of Sciences, 162 00 Prague, Czech Republic}
	
	\author{\mbox{M. Šindler}}
	\affiliation{Institute of Physics, Czech Academy of Sciences, 162 00 Prague, Czech Republic}
	
	\author{\mbox{T. Helm}}
	\affiliation{Hochfeld-Magnetlabor Dresden (HLD-EMFL), Helmholtz-Zentrum Dresden-Rossendorf, 01328 Dresden, Germany}
	
	\author{\mbox{S. Fukami}}
	\affiliation{Research Institute of Electrical Communication, Tohoku University, Sendai 980-8577 Japan}
	
	\author{\mbox{F. Krizek}}
	\affiliation{Institute of Physics, Czech Academy of Sciences, 162 00 Prague, Czech Republic}
	
	\author{\mbox{D. Kriegner}}
	\affiliation{Institute of Physics, Czech Academy of Sciences, 162 00 Prague, Czech Republic}
	\email{kriegner@fzu.cz}
	
	\date{\today}

	\begin{abstract}
		We report superconductivity in epitaxial PtSb(0001) thin films grown on \ce{SrF2}(111).
		Electrical transport measurements reveal a superconducting transition at $T_{\mathrm c}=\SI{1.72}{\kelvin}$.       
        The field-induced broadening of the transition and the presence of finite upper critical fields are consistent with type-II superconductivity.
		We determine the resistively defined upper critical fields for magnetic fields applied perpendicular and parallel to the film plane and parameterize their temperature dependence using an anisotropic Ginzburg--Landau approach.
		For the thickest film ($d=\SI{50}{\nano\meter}$), this yields coherence lengths of $\xi_{ab}\approx\SI{55}{\nano\meter}$ and $\xi_c\approx\SI{14}{\nano\meter}$.
		Current--voltage characteristics show sizeable critical currents, with a critical current density reaching $J_{\mathrm c}\approx\SI{6e4}{A/cm^2}$ at \SI{0.5}{\kelvin}.
        These results establish epitaxial PtSb as a superconducting thin-film platform compatible with lattice-matched heterostructures in the NiAs-type materials family. 
	\end{abstract}

	\maketitle

\section{\label{sec:introduction}Introduction}

Epitaxial superconducting thin films provide a materials basis for both quantum and cryogenic electronics and for controlled proximity heterostructures, where interfacial quality, crystallographic registry, and symmetry matching can be decisive~\cite{krantz2019, kjaergaard2020}.
In comparison to polycrystalline layers, single-crystalline films enable reproducible nanofabrication, well-defined anisotropy studies, and systematic interface engineering in superconductor/metal and superconductor/magnet hybrid heterostructures.
More broadly, proximity-coupled superconducting heterostructures constitute a central route toward superconducting spintronics and related device concepts~\cite{linder2015,buzdin2005,bergeret2005,eschrig2011}.

In this context, the NiAs-type structure (space group $P6_3/mmc$, No.~194) is an appealing ``modular'' materials family: it hosts multiple superconductors as well as a wide range of magnetic and semiconducting compounds with closely matched lattice parameters, enabling epitaxial heterostructures~\cite{burrows2019,bommanaboyena2025}.
Besides PtSb, bulk superconductivity has been reported in several $P6_3/mmc$ intermetallics such as PdSb, PdTe, NiBi, PtBi, PtSn, and RhBi~\cite{matthias1953,Matthias1953_2,Matthias1963,Chen2022,Karki2012,Roberts1976,Hamilton1965}.
At the same time, closely related $P6_3/mmc$ magnets provide complementary building blocks for heterostructures, such as ferromagnets MnSb and CrTe~\cite{takei1963,takei1966}.
This breadth motivates establishing epitaxial thin-film superconductors within this structure family as a basis for lattice-matched superconducting/magnetic stacks.

A particularly timely development in the same structural ecosystem is the discovery of altermagnetism, a form of collinear compensated magnetic order that yields momentum-dependent spin splitting without net magnetization~\cite{smejkal2022,smejkal2022a}.
Experimental signatures of altermagnetism, including direct evidence of altermagnetic band splitting, have been reported for NiAs-type MnTe~\cite{gonzalezbetancourt2023,krempasky2024,lee2024,osumi2024} and CrSb~\cite{reimers2024a,zhou2025}.
Thin-film growth of MnTe and CrSb has also been demonstrated, establishing both compounds as materials building blocks for epitaxial heterostructures~\cite{kriegner2016,reimers2024a,bommanaboyena2025,tseng2025}.
Several theoretical works have proposed that superconductor/altermagnet hybrids can host proximity-driven phenomena beyond conventional superconductor/ferromagnet structures, including characteristic modifications of Andreev reflection and phase-shifted Andreev bound states~\cite{papaj2023,sun2023,beenakker2023,niu2024}.
Related predictions include phase shifts and anomalous Josephson effects, finite-momentum Cooper pairing, and superconducting diode or other nonreciprocal responses~\cite{ouassou2023,lu2024,zhang2024,banerjee2024,cheng2024,chakraborty2024,chakraborty2025,chakraborty2025a,giil2024,giil2024a,sumita2025,boruah2025,mazin2025,sukhachov2025}.
This provides additional motivation to develop superconducting thin films compatible with $P6_3/mmc$ altermagnets.

Here we report superconductivity in single-crystalline PtSb(0001) epitaxial thin films grown on \ce{SrF2}(111).
Recent bulk work by Hirai \textit{et al.} on the high-entropy antimonide series (RuRhPdIr)$_{1-x}$Pt$_x$Sb highlighted composition- and disorder-tunable superconductivity, including an enhanced $H_{c2}$ in the strongly disordered regime and a PtSb end member consistent with earlier reports~\cite{Hirai2024}.
Superconductivity in epitaxial PtSb thin films has, to the best of our knowledge, remained largely unexplored.
We combine structural characterization with transport measurements to determine the superconducting transition and quantify the anisotropic upper critical fields and critical current densities.
By establishing epitaxial superconducting PtSb thin films within the NiAs-type family, our work provides a materials basis for future lattice-matched proximity heterostructures, including superconductor/magnet stacks involving NiAs-type altermagnets.

\section{\label{sec:methods}Experimental methods}

Epitaxial PtSb(0001) thin films were grown by dc magnetron co-sputtering from elemental Pt and Sb targets (99.99\% purity) in a confocal sputtering system.
The chamber base pressure was below $3\times10^{-8}$\,mbar and sputtering was performed in high-purity Ar (99.999\%) at a working pressure of $6.5\times10^{-3}$\,mbar.
The individual sputter fluxes were monitored by an in situ quartz crystal microbalance and adjusted to obtain stoichiometric PtSb.
SrF$_2$(111) substrates (Crystec GmbH) were chosen because they are chemically inert and electrically insulating, and because their hexagonal in-plane surface lattice constant ($a_{\parallel}=a_{\mathrm{SrF_2}}/\sqrt{2}$) provides a close lattice match to PtSb(0001).
The substrates were held at \SI{400}{\celsius} during deposition.
Further details about the growth can be found in Ref.~\cite{bommanaboyena2025}.

Crystallographic phase purity and out-of-plane orientation were characterized by X-ray diffraction (XRD) symmetric radial scans.
Surface morphology was examined by tapping-mode atomic force microscopy (AFM).
Cross-sectional microstructure and interface quality were investigated by high-angle annular dark-field imaging with a probe-corrected scanning transmission electron microscope (HAADF-STEM) on FIB-prepared lamellae.

For transport measurements, the films were patterned into Hall-bar devices by optical lithography followed by Ar$^+$ ion milling and measured in a four-probe configuration, with typical channel widths $w\sim 8$--\SI{18}{\micro\meter} and voltage-probe separations $L\sim 40$--\SI{360}{\micro\meter}; film thicknesses were in the range $d=\SI{15}{\nano\meter}$--\SI{50}{\nano\meter}.
The superconducting transition temperature and upper critical fields were determined from resistance measurements as a function of temperature and magnetic field for field orientations perpendicular and parallel to the film plane (criteria specified in the main text).
Measurements above $\sim0.5$\,K were carried out in a Quantum Design PPMS with a $^3$He insert, while lower-temperature data were acquired in a dilution refrigerator using low-frequency lock-in techniques.
The differential resistance $\mathrm{d}V/\mathrm{d}I$ was obtained by numerically differentiating the measured $V(I)$ characteristics as a function of dc bias current and used to determine the critical current density.

\section{\label{sec:results}Results}

\subsection{\label{sec:structure}Epitaxy and structural characterization}

\begin{figure}[tbp]
	\centering
	\includegraphics[width=1\columnwidth, keepaspectratio]{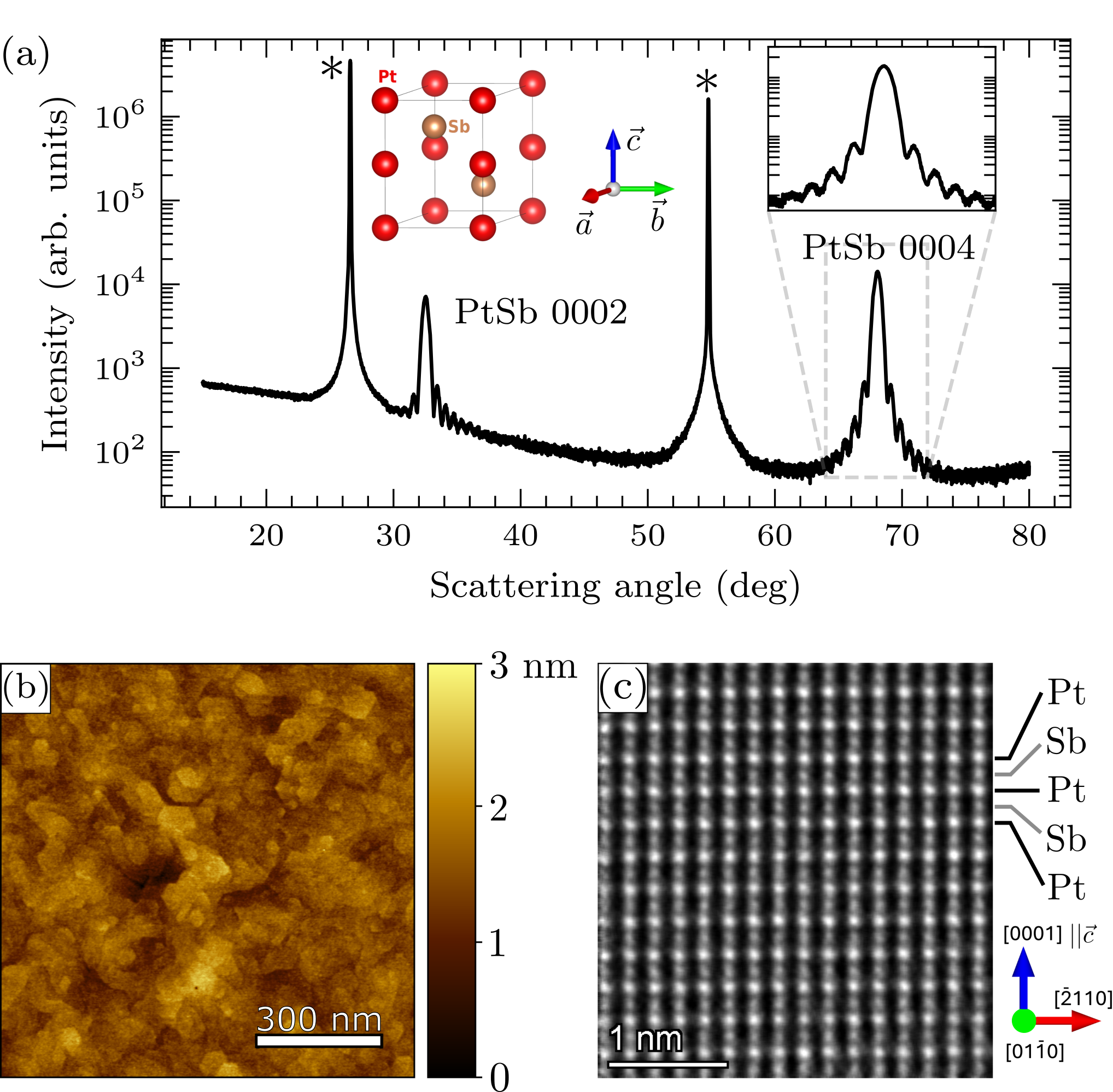}
	\caption{\label{fig:1} Structural characterization of epitaxial PtSb(0001) thin films on \ce{SrF2}(111).
		(a) Symmetric XRD radial scan of a \SI{15}{\nano\meter} PtSb(0001) film showing only PtSb 000$\ell$ reflections together with the substrate peaks marked by a star; Laue thickness fringes indicate uniform thickness and high crystalline coherence. An inset shows the unit cell structure of PtSb \cite{Momma2011}.
		(b) Tapping-mode AFM topography of a \SI{30}{\nano\meter} PtSb(0001) film, revealing $\sim$10--\SI{100}{\nano\meter} terraces separated by bi-atomic steps.
		(c) Cross-sectional HAADF-STEM image along the $[01\bar{1}0]$ zone axis, resolving the layered stacking of Pt and Sb atomic planes.}
\end{figure}

Figure~\ref{fig:1} summarizes the structural characterization of epitaxial PtSb(0001) thin films grown on SrF$_2$(111).
PtSb crystallizes in the hexagonal NiAs-type structure with reported bulk lattice parameters $a=\SI{4.13}{\angstrom}$ and $c=\SI{5.48}{\angstrom}$~\cite{Kjekshus1969,Zhuravlev1962,Ellner2004}.
The symmetric X-ray diffraction (XRD) radial scan in Fig.~\ref{fig:1}(a) shows only PtSb 000$\ell$ Bragg reflections together with the substrate peaks, confirming phase purity and the (0001) growth plane orientation.
Pronounced Laue thickness fringes around the PtSb peaks further indicate a uniform thickness and high crystalline coherence.
From the out-of-plane peak positions we extract a $c$-axis lattice parameter of $c=\SI{5.5}{\angstrom}$, in good agreement with literature.
The in-plane lattice parameter was determined independently from reciprocal space maps (not shown) and is found to be $a=\SI{4.12}{\angstrom}$.
This corresponds to an in-plane lattice mismatch of order $<1\%$ relative to the SrF$_2$(111) surface lattice constant, consistent with epitaxial growth.
The surface morphology, characterized by tapping-mode atomic force microscopy (AFM) [Fig.~\ref{fig:1}(b)], exhibits extended terraces with lateral sizes on the order of $10$--$100$~nm separated by bi-atomic steps, consistent with step-flow growth on a hexagonal lattice.
The root-mean-square surface roughness is $\SI{3}{\angstrom}$, in good agreement with the value obtained from X-ray reflectivity (XRR, not shown), and the peak-to-valley height variation amounts to $\approx\SI{2.5}{\nano\meter}$.
Finally, the HAADF-STEM image in Fig.~\ref{fig:1}(c), taken along the $[01\bar{1}0]$ zone axis, resolves the layered stacking of alternating Pt- and Sb-containing atomic planes, corroborating the single-crystalline nature of the films and a sharp film/substrate interface.
Together, these measurements establish epitaxial PtSb(0001) thin films with high structural quality as a basis for the superconducting transport experiments discussed below.

\subsection{\label{sec:transition}Superconducting transition}

\begin{figure}[tbp]
	\centering
	\includegraphics[width=1\columnwidth]{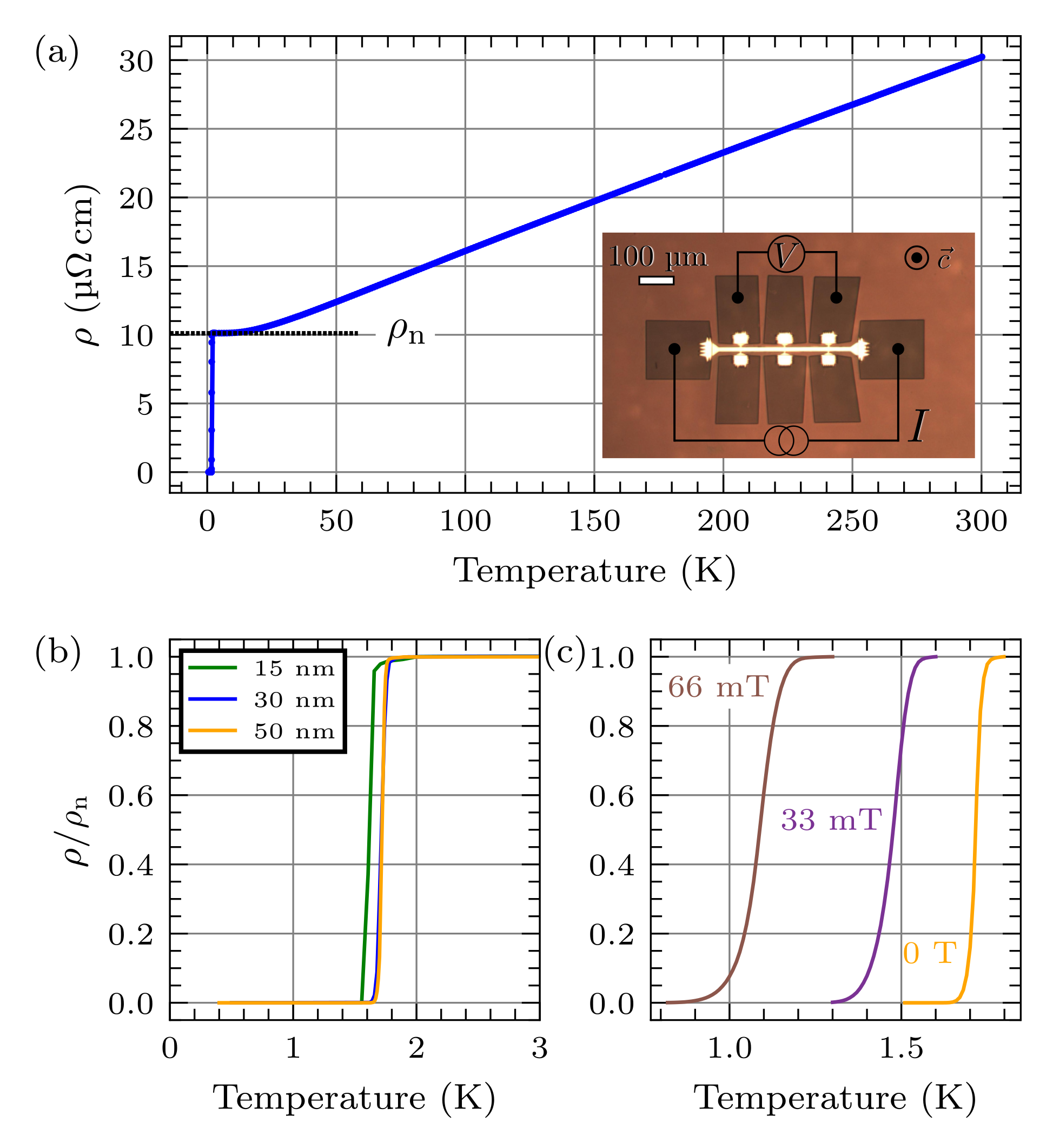}
	\caption{\label{fig:2} Superconducting transition in PtSb thin films.
		(a) Resistivity $\rho(T)$ at zero magnetic field for a \SI{30}{\nano\meter} thick PtSb epitaxial film. An inset shows an optical microscope image of the used device structure.
		(b) Superconducting transitions in the normalized resistivity for films with thicknesses of \SI{15}{\nano\meter}, \SI{30}{\nano\meter} and \SI{50}{\nano\meter} (the \SI{30}{\nano\meter} and \SI{50}{\nano\meter} traces largely overlap).
		(c) Normalized resistivity $\rho(T)/\rho_{\mathrm n}$ of a \SI{50}{\nano\meter} film at various magnetic fields applied perpendicular to the film plane, where $\rho_{\mathrm n}$ denotes the normal-state resistivity (see main text).}
\end{figure}

Figure~\ref{fig:2}(a) shows the temperature dependence of the resistivity $\rho(T)$ of a representative PtSb device at zero applied magnetic field.
In the normal state, the resistivity at $T=\SI{2}{\kelvin}$ is $\rho_{\mathrm n}=\SI{10.1}{\micro\ohm\centi\meter}$, corresponding to a residual resistivity ratio $\mathrm{RRR}=\rho(\SI{300}{\kelvin})/\rho(\SI{2}{\kelvin})\approx 3$.
We define the superconducting transition temperature $T_{\mathrm c}$ as the temperature where $\rho(T)$ drops to $50\%$ of $\rho_{\mathrm n}$, yielding $T_{\mathrm c}=\SI{1.72}{\kelvin}$ for the device shown.
The transition width $\Delta T_{\mathrm c}$, defined as the temperature interval between $0.9\rho_{\mathrm n}$ and $0.1\rho_{\mathrm n}$, is $\Delta T_{\mathrm c}=\SI{0.05}{\kelvin}$.

Figure~\ref{fig:2}(b) compares the superconducting transitions for films with thicknesses $d=\SI{15}{\nano\meter}$, \SI{30}{\nano\meter}, and \SI{50}{\nano\meter}.
The transitions of the \SI{30}{\nano\meter} and \SI{50}{\nano\meter} films largely overlap, while the \SI{15}{\nano\meter} film exhibits a slightly reduced $T_{\mathrm c}$.
Such small variations may reflect modest changes in disorder, strain, or finite-size effects between films, and will not be further analyzed here.

To examine the response to magnetic field, Fig.~\ref{fig:2}(c) shows the normalized resistivity $\rho(T)/\rho_{\mathrm n}$ of a \SI{50}{\nano\meter} film in magnetic fields applied perpendicular to the film plane.
With increasing field, the transition shifts to lower temperature and broadens systematically, as commonly observed in thin-film superconductors; a more detailed discussion of the field-induced broadening is deferred to the discussion of the critical-field data below.

\subsection{\label{sec:hc2}Upper critical fields and coherence lengths}

\begin{figure}[tbp]
	\centering
	\includegraphics[width=1\columnwidth]{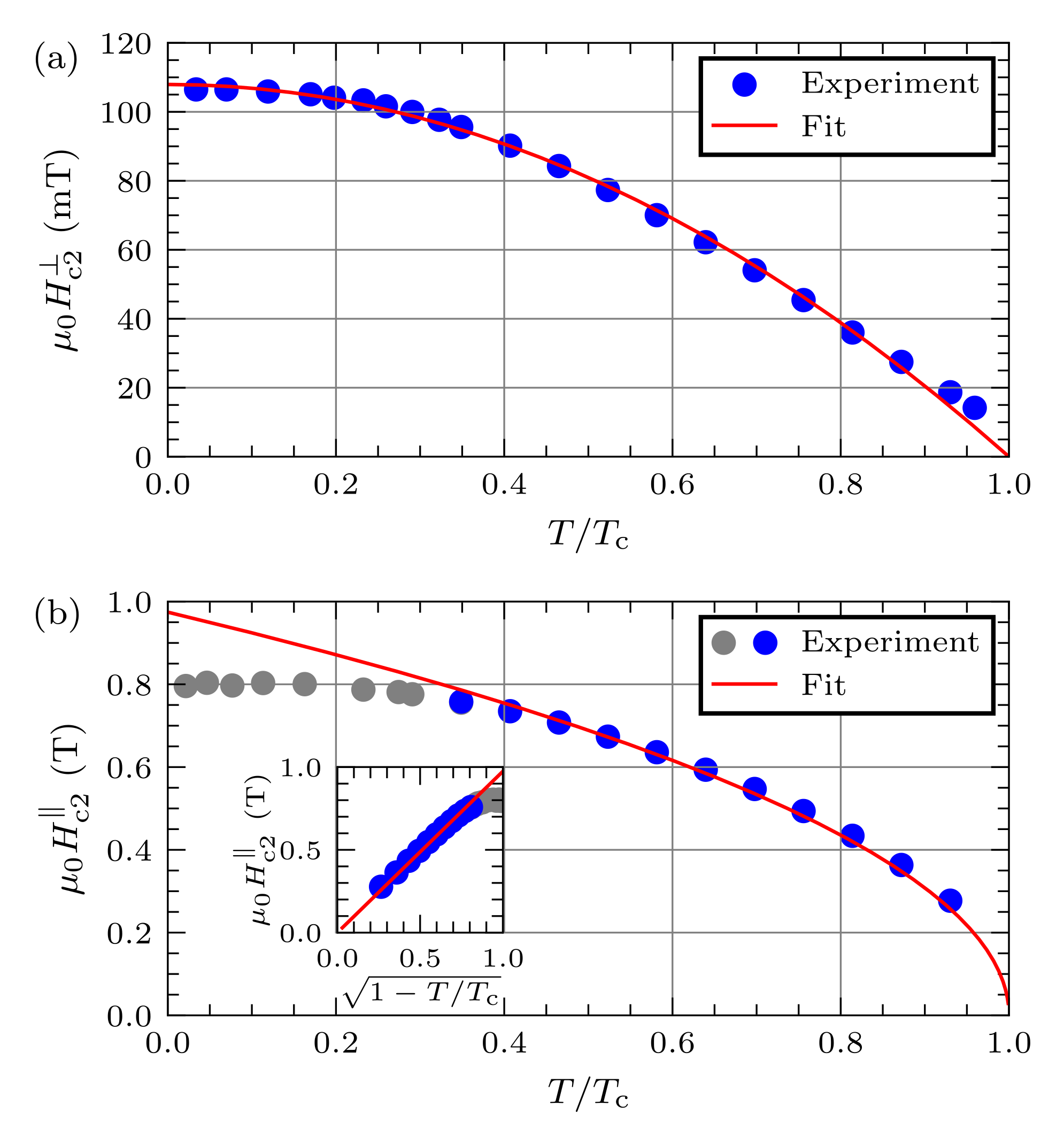}
	\caption{\label{fig:3} Upper critical fields for magnetic fields applied perpendicular and parallel to the film plane.
		(a) Perpendicular upper critical field $\mu_0H_{\mathrm c2}^{\perp}$ versus temperature measured in a \SI{50}{\nano\meter} thick PtSb epitaxial film; the solid line is a fit using Eq.~(\ref{eq:perpcriticalfields}).
		(b) Parallel upper critical field $\mu_0H_{\mathrm c2}^{\parallel}$ versus temperature measured in a \SI{30}{\nano\meter} thick film; the solid line is a fit using Eq.~(\ref{eq:parallelcriticalfields}) to the data points shown in blue.
		The inset shows the same data plotted versus $\sqrt{1-\sfrac{T}{T_{\mathrm c}}}$, emphasizing the approximately linear behavior close to $T_{\mathrm c}$.}
\end{figure}

To quantify the superconducting anisotropy, we determined the resistively defined upper critical field $H_{\mathrm c2}$ for magnetic fields applied perpendicular ($\perp$) and parallel ($\parallel$) to the film plane.
The temperature dependences $\mu_0H_{\mathrm c2}^{\perp}(T)$ and $\mu_0H_{\mathrm c2}^{\parallel}(T)$ are summarized in Fig.~\ref{fig:3}.
The critical fields were extracted from magnetic-field sweeps using a resistive criterion of $0.9\rho_{\mathrm n}$, where $\rho_{\mathrm n}$ denotes the normal-state resistivity.
For each temperature, the reported values correspond to the mean of the absolute critical fields obtained from forward and backward sweeps in both field polarities.

To parameterize the data we employ standard anisotropic Ginzburg--Landau (GL) relations~\cite{tinkham2004introduction,Harper1968,Xing2017}.
For $\mu_0H\perp$ film plane we use
\begin{equation}
	\mu_0H_{\mathrm c2}^{\perp}(T)=\frac{\Phi_0}{2\pi\xi_{ab}^2}\left[1-\left(\frac{T}{T_\mathrm c}\right)^2\right],
	\label{eq:perpcriticalfields}
\end{equation}
where $\Phi_0=h/2e$ is the magnetic flux quantum and $\xi_{ab}$ denotes the in-plane coherence length.
For $\mu_0H\parallel$ film plane we use
\begin{equation}
	\mu_0H_{\mathrm c2}^{\parallel}(T)=\frac{\Phi_0}{2\pi\xi_{ab}\xi_c}\sqrt{1-\frac{T}{T_\mathrm c}},
	\label{eq:parallelcriticalfields}
\end{equation}
with $\xi_c$ the out-of-plane coherence length and $\xi_{ab}$ is taken from the $\mu_0H\perp$ film plane analysis.
The resulting critical fields and coherence lengths for different film thicknesses are summarized in Table~\ref{tab:coherence}.

As shown in Fig.~\ref{fig:3}(a), $\mu_0H_{\mathrm c2}^{\perp}(T)$ is well described by Eq.~(\ref{eq:perpcriticalfields}) over the measured temperature range.
For $\mu_0H\parallel$ film plane [Fig.~\ref{fig:3}(b)], $\mu_0H_{\mathrm c2}^{\parallel}(T)$ follows the expected near-$T_{\mathrm c}$ trend but exhibits a weaker temperature dependence at low $T$.
To extract fit parameters representative of the near-$T_{\mathrm c}$ regime, we therefore restrict the fit of $\mu_0H_{\mathrm c2}^{\parallel}(T)$ to the temperature interval where the linearized plot in the inset of Fig.~\ref{fig:3}(b) shows an approximately linear dependence on $\sqrt{1-T/T_{\mathrm c}}$, as expected from Eq.~(\ref{eq:parallelcriticalfields}).
In this regime, $H_{\mathrm c2}$ is approximately linear in $(T_{\mathrm c}-T)$, consistent with the behavior discussed for thin-film superconductors by Drew \textit{et al.}~\cite{drew2009}.
While this procedure improves the fit quality within the selected range, it increases the deviation between the extrapolated fit and the lowest-temperature data to about $20\%$ for the present dataset.
Accordingly, Table~\ref{tab:coherence} reports $\xi_c$ values obtained both from the restricted-range fits and from an estimate based on the lowest-temperature data (see caption).
Overall, the field-orientation dependence and finite values of $H_{\mathrm c2}$ are consistent with anisotropic type-II superconductivity in PtSb, in line with magnetization data reported for related antimonides including the PtSb end member~\cite{Hirai2024}.
Using the values in Table~\ref{tab:coherence}, the film thicknesses satisfy $d\gtrsim \xi_c$ for all samples, indicating that the films are not in the strict two-dimensional limit.
Furthermore, the measured $\mu_0H_{\mathrm c2}^{\parallel}$ values are well below the weak-coupling Pauli paramagnetic limit $\mu_0H_{\mathrm P}\approx 1.84\,T_{\mathrm c}$~\cite{clogston1962,chandrasekhar1962}, which for $T_{\mathrm c}=\SI{1.72}{\kelvin}$ yields $\mu_0H_{\mathrm P}\approx\SI{3.2}{\tesla}$.
This suggests that orbital pair breaking provides the dominant limitation in the present field range.
Within anisotropic Ginzburg--Landau theory the superconducting anisotropy can be expressed as $\gamma=\mu_0H_{\mathrm c2}^{\parallel}/\mu_0H_{\mathrm c2}^{\perp}=\xi_{ab}/\xi_c$~\cite{tinkham2004introduction}.
The resulting $\gamma$ values (Table~\ref{tab:coherence}) increase systematically with decreasing film thickness; using the low-temperature estimates for $\mu_0H_{\mathrm c2}^{\parallel}$ yields slightly different absolute values but does not change this trend.
We note that, owing to the sharp resistive transition, varying the resistive criterion within reasonable bounds has only a minor influence on the extracted $\mu_0H_{\mathrm c2}$ values.

\begin{table}[tbp]
	\centering
\centering
\setlength{\tabcolsep}{8pt}
\renewcommand{\arraystretch}{1.2}
\begin{tabular}{lccccc}
\toprule
\makecell{$d$ \\ (nm)} &
\makecell{$\mu_0 H_{\mathrm c2}^{\perp}$ \\ (mT)} &
\makecell{$\xi_{ab}$ \\ (nm)} &
\makecell{$\mu_0 H_{\mathrm c2}^{\parallel}$ \\ (mT)} &
\makecell{$\xi_c$ \\ (nm)} &
$\gamma$ \\
\midrule
15 & 108 & 55 & 2169, 2500 & 3.0, 2.4 & 23.2 \\
30 & 74  & 67 & 796, 974   & 6.5, 5.0 & 13.2 \\
50 & 108 & 55 & 425, 535   & 14, 11   & 5 \\
\bottomrule
\end{tabular}
\caption{\label{tab:coherence} Upper critical fields, coherence lengths, and anisotropy parameter for different film thicknesses.
	The perpendicular upper critical field $\mu_0H_{\mathrm c2}^{\perp}$ and the in-plane coherence length $\xi_{ab}$ were obtained from fits using Eq.~(\ref{eq:perpcriticalfields}).
	For the parallel field, the two values of $\mu_0H_{\mathrm c2}^{\parallel}$ (and the corresponding $\xi_c$) correspond to an estimate based on the lowest-temperature data and to the $\SI{0}{\kelvin}$ value extrapolated from fits using Eq.~(\ref{eq:parallelcriticalfields}), respectively.
	The anisotropy parameter $\gamma=\mu_0H_{\mathrm c2}^{\parallel}/\mu_0H_{\mathrm c2}^{\perp}=\xi_{ab}/\xi_c$~\cite{tinkham2004introduction} is calculated from the fit-derived values (second entries).}
\end{table}

\subsection{\label{sec:jc}Critical current density}

\begin{figure}[h!]
	\centering
	\includegraphics[width=1\columnwidth]{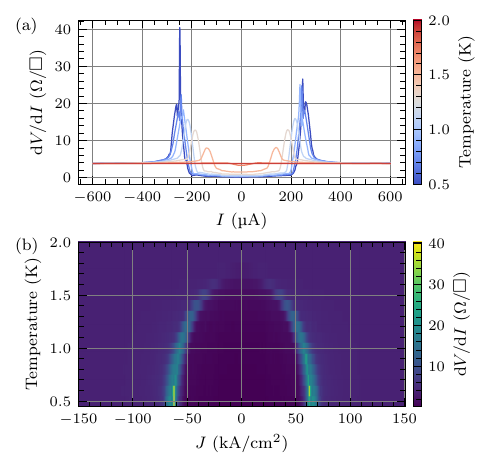}
    \caption{\label{fig:IV} Current-driven transport in PtSb.
	(a) Differential resistance $\mathrm{d}V/\mathrm{d}I$, obtained by numerical differentiation of measured $V(I)$ characteristics, shown as a function of current for a \SI{50}{\nano\meter} thick PtSb film at temperatures from \SI{0.5}{\kelvin} to \SI{1.9}{\kelvin} in \SI{0.2}{\kelvin} steps.
    Here $\mathrm{d}V/\mathrm{d}I$ is converted to a differential sheet resistance (in $\Omega/\square$) to normalize out the Hall-bar geometry.
	(b) Color map of $\mathrm{d}V/\mathrm{d}I$ as a function of current density $J$ and temperature.
	The critical current density $J_{\mathrm c}$ is extracted from the position of the maximum in $\mathrm{d}V/\mathrm{d}I$ (see text) and reaches $\approx \SI{60}{kA/cm^2}$ at \SI{0.5}{\kelvin}.}
\end{figure}

Figure~\ref{fig:IV} summarizes the current-driven transport of PtSb in the superconducting state.
Current--voltage characteristics $V(I)$ were recorded at fixed temperature, and the differential resistance $\mathrm{d}V/\mathrm{d}I$ shown in Fig.~\ref{fig:IV} was obtained by numerical differentiation.
Figure~\ref{fig:IV}(a) displays $\mathrm{d}V/\mathrm{d}I(I)$ for a \SI{50}{\nano\meter} thick film measured at temperatures from \SI{0.5}{\kelvin} to \SI{1.9}{\kelvin}.
At low temperatures, $\mathrm{d}V/\mathrm{d}I$ exhibits a sharp peak at a characteristic bias current, which shifts to lower current and broadens upon increasing temperature, consistent with a progressive suppression of superconductivity.

We define the critical current $I_{\mathrm c}$ as the current at which $\mathrm{d}V/\mathrm{d}I$ reaches its maximum, corresponding to the steepest point of the measured $V(I)$ characteristic for a given temperature.
The critical current density is then obtained as $J_{\mathrm c}=I_{\mathrm c}/(wd)$, where $w$ and $d$ are the device width and film thickness, respectively.
The resulting evolution of $\mathrm{d}V/\mathrm{d}I$ as a function of current density $J$ and temperature is summarized in the color map in Fig.~\ref{fig:IV}(b).
Using this definition, we obtain a critical current density of $J_{\mathrm c} \approx \SI{60}{kA/cm^2}$ at $T=\SI{0.5}{\kelvin}$ for the \SI{50}{\nano\meter} film.
All $R(T)$ and $R(H)$ measurements discussed above were performed at substantially smaller excitation current densities (typically $\sim\SI{2.5}{kA/cm^2}$), ensuring that the extracted transition temperatures and upper critical fields are not affected by bias-current-induced suppression.

\section{\label{sec:discussion}Discussion and outlook}

The data in Figs.~\ref{fig:1}--\ref{fig:IV} establish epitaxial PtSb(0001) thin films on SrF$_2$(111) as a reproducible superconducting thin-film platform.
Structural characterization (Fig.~\ref{fig:1}) evidences single-crystalline growth with low surface roughness and lattice parameters close to bulk values, providing a well-defined basis for transport experiments.
The resistive transitions (Fig.~\ref{fig:2}) are sharp and consistent across multiple thicknesses, with only a modest reduction of $T_{\mathrm c}$ in the thinnest film.
Such small variations are common in superconducting thin films and can reflect subtle differences in disorder, strain, or finite-size effects.
The field-broadened resistive transitions observed upon applying a perpendicular magnetic field [Fig.~\ref{fig:2}(c)] are commonly found in superconductors where vortices, thermal fluctuations, and dissipative vortex motion can contribute to finite resistance below the upper critical field~\cite{Zeinali2016,Marchenko1993,Wang2010}.
The observation of finite upper critical fields is consistent with type-II superconductivity.
In support of this classification, magnetization data reported for the related antimonide series (RuRhPdIr)$_{1-x}$Pt$_x$Sb, including the PtSb end member, show the characteristic behavior expected for a type-II superconductor~\cite{Hirai2024}.
While we do not pursue a detailed analysis of the transition broadening here, the systematic shift and broadening with field are consistent with the overall picture of anisotropic superconductivity in these films.

Upper critical fields measured for out-of-plane and in-plane field orientations (Fig.~\ref{fig:3}) reveal pronounced anisotropy that increases with decreasing thickness.
Parameterizing the data with anisotropic Ginzburg--Landau relations yields in-plane and out-of-plane coherence lengths summarized in Table~\ref{tab:coherence}.
For $\mu_0H\parallel$ film plane, the data follow the expected trend close to $T_{\mathrm c}$ but exhibit a weaker temperature dependence at low temperatures.
Similar low-temperature deviations from simple near-$T_{\mathrm c}$ parameterizations have been discussed in other thin-film superconductors~\cite{Wang2002,Broussard2017,Coffey2010,Xing2017}, and may reflect the limited validity of the employed functional form away from the transition and/or additional physics not captured by the minimal model.
Importantly, the resulting coherence lengths satisfy $d\gtrsim \xi_c$ for all samples, indicating that the films are not in the strict two-dimensional limit.

Placing the extracted parameters into the broader context of NiAs-type superconductors, the in-plane coherence length of the thickest PtSb film is comparable to values reported for bulk PdSb ($\xi_{ab}\approx\SI{61}{\nano\meter}$)~\cite{Chen2022} and bulk PdTe ($\xi_{ab}\approx\SI{48}{\nano\meter}$)~\cite{Karki2012}.
Likewise, the superconducting transition temperatures of these NiAs-type intermetallics fall in a similar low-$T_{\mathrm c}$ range~\cite{matthias1953,Matthias1953_2,Matthias1963}.
Together, this comparison supports that epitaxial PtSb thin films realize superconductivity with characteristic length scales typical for this structure family, while offering the key advantage of compatibility with lattice-matched epitaxial heterostructures.

Current-biased measurements (Fig.~\ref{fig:IV}) demonstrate that PtSb thin films sustain sizable critical current densities at low temperature.
Since, to the best of our knowledge, critical current densities have not been systematically reported for NiAs-type superconductors, we place our values into the broader context of established thin-film superconductors.
The critical current density extracted here, $J_{\mathrm c}\approx \SI{6e4}{A/cm^2}$ at \SI{0.5}{\kelvin}, is of the same order as transport critical currents commonly reported for conventional thin-film superconductors used in quantum and cryogenic electronics, such as Nb, NbN, TaN, and Al-based bridges and nanowires, noting that reported values depend strongly on geometry, disorder, and the chosen extraction criterion~\cite{ilin2014,mercereau1962,miller1993,morgan-wall2015}.
These $J_{\mathrm c}$ values show that PtSb thin films can sustain sizeable supercurrents in lithographically defined devices.
Together with the anisotropic $H_{\mathrm c2}$ behavior and the structural quality, this establishes epitaxial PtSb(0001) as a reproducible superconducting thin-film material within the NiAs-type family.

\section{\label{sec:conclusion}Conclusion}

We have established superconductivity in single-crystalline PtSb(0001) epitaxial thin films grown on SrF$_2$(111) and quantified key transport characteristics relevant for thin-film and heterostructure implementations.
From resistive measurements we determine the superconducting transition, the anisotropic upper critical fields, coherence lengths, and critical current densities, and find parameters consistent with superconductivity in the NiAs-type materials family.
These results position epitaxial PtSb as a promising superconducting thin-film material and motivate future lattice-matched proximity heterostructures within the NiAs-type platform, including superconductor/magnet stacks involving recently identified altermagnets.

\section{\label{sec:acknowledgments}Acknowledgments}

We acknowledge financial support by the Czech Science Foundation (Grant No. 22-22000M), Lumina Quaeruntur fellowships LQ100102201 and LQ100102602 of the Czech Academy of Sciences, Czech Ministry of Education, Youth and Sports (MEYS) grants LM2023051 and CZ.02.01.01/00/22\_008/0004594 and ERC Advanced Grant No. 101095925. We are grateful to Zbyněk~Šobáň from the Institute of Physics, Czech Academy of Sciences for technical support and assistance. Experiments were performed in MGML (mgml.eu), which is supported within the program of Czech Research Infrastructures (project no. LM2023065). We acknowledge CzechNanoLab Research Infrastructure supported by MEYS CR (LM2023051).

\bibliography{references}

\end{document}